\documentclass{intech08a}

\newcommand{\apj}{ApJ}
\newcommand{\apjs}{ApJS}
\newcommand{\aj}{AJ}
\newcommand{\aap}{A\&A}
\newcommand{\aapr}{A\&AR}
\newcommand{\mnras}{MNRAS}
\newcommand{\pasp}{PASP}
\newcommand{\nat}{Nature}
\newcommand{\apjl}{ApJL}
\newcommand{\araa}{ARofA\&A}
\newcommand{\nar}{NewAR}
\newcommand{\memsai}{MdSAI}

\booktitle{Will-be-set-by-IN-TECH}%
\headertitle{Direct Imaging of Extra-solar Planets - Homogeneous Comparison of Detected
Planets and Candidates}

\chaptertitle{Direct Imaging of Extra-solar Planets -\\
\vspace{-1mm}
 Homogeneous Comparison of Detected \\
Planets and Candidates} % You know your chapter title?

\authors{Ralph Neuh\"auser and Tobias Schmidt}
\affiliation{Astrophysical Institute and University Observatory, Friedrich Schiller University Jena, Schillerg\"asschen 2-3,
07745 Jena}
\country{Germany}

\begin{document}

\maketitle

%FOR AUTHOR: E-mail addresses are not allowed in manuscript

\section{Introduction}

Planets orbiting stars other than the Sun are called {\em extra-solar planets}
or {\em exo-planets}. Since about 1989, several hundred such objects were detected
using various techniques. The first companions with planetary masses orbiting
another star were found around a pulsar, a fast rotating neutron star,
by variations of the otherwise very stable radio pulses \citep{1992Natur.355..145W}.
With the radial velocity technique, one can detect the
motion of the star in radial direction towards and away from us due to the fact
that both a star and a planet orbit their common center of mass
(first successfully done on HD 114762 and 51 Peg, \citet{1989Natur.339...38L} and \citet{1995Natur.378..355M},
respectively). This one-dimensional
technique yields the lower mass limit $m \cdot \sin i$ of the companion mass $m$ due to the
unknown orbit inclination $i$, so that planets detected only by the radial velocity
method are {\em planet candidates}, they could also be higher mass brown dwarfs or low-mass stars.
Several hundred planet candidates were detected with this method.
The reflex motion of the star in the two other dimensions due to the orbiting companion
can be measured with the astrometry method (e.g. \citet{2002ApJ...581L.115B}),
but no new planets were found with this technique so far.
If the orbital plane of a planet is in the line of sight towards the Earth, then the planet will
move in front of the star once per orbital period, which can be observed as transit, i.e. as
small drop in the brightness of the star. This determines the inclination $i$ and, for
radial velocity planet candidates, can confirm candidates to be true planets.
The transit method could confirm almost one dozen planet candidates previously
detected by the radial velocity method (the first was HD 209458 b, \citet{2000ApJ...529L..45C});
in addition, more than 100 planets were
originally discovered as candidates by the transit method and then confirmed by radial velocity.
All the techniques mentioned above are {\em in}direct techniques, i.e. they all observe
only the star, but not the planet or planet candidate, i.e. it is never known which photons
were emitted by the star (most) and which by the planets (negligibly few).
This is different only in the direct imaging technique, which we discuss below in detail.
For number of planets, lists, properties, and references, see \citet{2011A&A...532A..79S}
with updates on www.exoplanet.eu or \citet{2011PASP..123..412W} with updates on www.exoplanets.org.

For all techniques, it is also relevant to define what is called a {\em planet}.
In the case of extra-solar planet, what is relevant is to define the upper mass limit
of planets and the distinction from brown dwarfs.
For the Solar System, the International Astronomical Union (IAU) has defined the lower mass
limit for planets:
{\em A celestial body that (a) is in orbit around the Sun,
(b) has sufficient mass for its self-gravity to overcome rigid body forces, so
that it assumes a hydrostatic equilibrium (nearly round) shape,
and (c) has cleared the neighbourhood around its orbit} (www.iau.org).

For the upper mass limit of planets, there are still several suggestions:
\begin{itemize}
\item The IAU {\em Working Group on Extrasolar Planets} has agreed on the following preliminary working definition:
{\em Objects with true masses below the limiting mass for thermonuclear fusion of deuterium (currently calculated
to be 13 Jupiter masses for objects of solar metallicity) that orbit stars or stellar remnants are planets (no matter
how they formed). The minimum mass/size required for an extra-solar object to be considered a planet should be the
same as that used in our Solar System. Sub-stellar objects with true masses above the limiting mass for thermonuclear
fusion of deuterium are brown dwarfs, no matter how they formed nor where they are located.
Free-floating objects in young star clusters with masses below the limiting mass for thermonuclear fusion of deuterium
are not planets, but are sub-brown dwarfs} (www.iau.org).
\item The mass (or $m \cdot \sin i$) distribution of sub-stellar companions is bi-modal,
which may indicate that the two populations formed differently;
the deviding mass can be used to define planets as those below the
so-called {\em brown dwarf desert}, which lies at around $\sim 25$~M$_{\rm Jup}$
\citep{2006ApJ...640.1051G,2010lyot.confE..11U,2010arXiv1012.1319S,2011A&A...532A..79S};
\citet{2011A&A...532A..79S} in their catalog on www.exoplanet.eu now include all those
companions with mass below $\sim 25$~M$_{\rm Jup}$ within a $1~\sigma$ error.
\item One can try to distinuish between planets and brown dwarfs by formation, e.g. that planets
are those formed in circumstellar disks with solid or fluid cores and brown dwarfs being those
formed star-like by direct gravitational contraction. In such a case, the mass ranges may overlap.
\end{itemize}
There is still no consensus on the definition of planets and their upper mass limit.
The second and third suggestions above, however, may be consistent with each other,
because the bi-modal distribution in masses may just be a consequence of different formation mechanism.
We will use $\sim 25$~M$_{\rm Jup}$ within a $1~\sigma$ error as upper mass limit for this paper.

For a direct detection of a planet close to its host star, one has to overcome
the large dynamical range problem (see Fig. 1 and 4): The planet is much fainter than its host star
and very close to its bright host star. Normal Jupiter-like planets
around low-mass stars ($\sim 0.1$~M$_{\odot}$) with one to few Gyr age are 6 orders of magnitude
fainter than their host stars \citep{1997ApJ...491..856B} - unless the planet would
have a large albedo and would be very close to the star
and, hence, would reflect a significant amount of star light, but then it is
too close to the star for direct detection.
Another exception are young planets, which are self-luminous due to ongoing contraction
and maybe accretion, so that they are only 2 to 4 orders of magnitude fainter
(for 13 to 1 Jup mass planets, respectively) than their (young) host stars,
again for $0.1$~M$_{\odot}$ stars (\citet{1997ApJ...491..856B},\citet{1998A&A...337..403B}).
Hence, direct imaging of planets is less difficult around young stars
with ages up to a few hundred Myr.

In this article, we will compile the planets and candidates imaged directly so far:
We will compile all their relevant properties to compare them in a homogeneous way,
i.e. to estimate their luminosities, temperatures, and masses homogeneously.
So far, the different teams, who have found and published the objects, use
different methods to estimate the companion mass, which is the most critical
parameter to decide about the nature of the object as either a planet or a brown dwarf.
We will then also discuss each object individually.

\section{Adaptive Optics observations to detect candidates}

Given the problem of dynamical range mentioned above, i.e.~that planets are much
fainter than stars and very close to stars, one has to use Adaptive Optics (AO)
imaging in the near-infrared JHKL bands (1 to 3.5 $\mu$m), in order to directly
detect a planet, i.e.~to resolve it from the star. The infrared (IR) is best,
because planets are cool and therefore brightest in the near- to mid-IR,
while normal stars are brighter in the optical than in the IR.
Two example images are given in Fig. 1.

Before any planets or planet candidates became detectable by ground-based AO observations,
brown dwarfs as companions to normal stars were detected, because brown dwarfs are
more massive and, hence, brighter, Gl 229 B being the first one
(\citet{1995Natur.378..463N}, \citet{1995Sci...270.1478O}).

We will now present briefly the different observational techniques.

In {\em normal near-IR imaging} observations, even without AO, one would also take many short
exposures in order not to saturate the bright host star (typically on the
order of one second), with some offset either after each image or after
about one minute (the time-scale after which the Earth atmosphere changes
significantly) - called jitter or dither pattern. One can then subtract each
image from the next or previous image (or a median of recent images) in
order to subtract the background, which is actually foreground emission from
Earth atmosphere etc. Then, one can add up or median all images,
a procedure called shift+add. Without AO and/or with exposure times much
longer than the correlation time of the atmosphere, such images will be far
from the diffraction limit. Objects like TWA 5 B (Fig. 1) or HR 7329 B, also
discussed below, were detected by this normal IR imaging with the 3.5m ESO NTT/SofI
(\citet{2000A&A...360L..39N},\citet{2001A&A...365..514G}).

In {\em speckle imaging}, also without AO, each single exposure should be as short as the
correlation time of the atmosphere (at the given wavelength), so that each image
can be diffraction-limited. Then, one also applies the shift+add technique.
A faint planet candidate near TWA 7 was detected in this way with the 3.5m ESO NTT/Sharp
\citep{2000A&A...354L...9N}, but later rejected by spectroscopy \citep{2002osp..conf..383N}.

In {\em Adaptive Optics} (AO) IR imaging, each single exposure should also be short
enough, in order not to saturate on the bright host star. If the host star image would
be saturated, one cannot measure well the position of the photocenter of its PSF,
so that the astrometric precision for the common proper motion test would be low.
One also applies the shift+add technique. Most planets and candidates imaged directly
were detected by normal AO imaging, see e.g. Fig. 1 (TWA 5), but are also limited regarding
the so-called {\em inner working angle}, i.e. the lowest possible separation
(e.g. the diffraction limit), at which a faint planet can be detected.
The diffraction limit ($\sim\,\lambda$/D) at D=8 to 10 meter telescopes
in the K-band ($\lambda\,=\,2.2\,\mu$m) is 0.045 to 0.057 arc sec; one cannot improve the
image quality (i.e. obtain a smaller diffraction limit) by always increasing the
telescope size because of the seeing, the turbulence in the Earth atmosphere,
hence AO corrections.
One can combine the advantages of speckle and AO, if the individual exposures are
very short and if one then saves all exposures (so-called {\em cube mode} at ESO VLT NACO AO instrument).

If the host star nor any other star nearby (in the isoplanatic patch) is bright enough
as AO guide star, then one can use a {\em Laser Guide Star}, as e.g. in the Keck AO observations
of Wolf 940 A and B, a planet candidate \citep{2009MNRAS.395.1237B}, see below.

One can also place the bright host star behind a coronagraph, so that the magnitude
limit will be larger, i.e. fainter companions would be detectable.
However, one then cannot measure the photocenter position of the host star,
so that the astrometric precision for the common proper motion test would be low.
One can use a {\em semi-transparent coronagraph}, so that both star and companion
are detected. We show an example in Fig. 1, the star $\epsilon$ Eri, where one close-in
planet may have been detected by radial velocity and/or astrometry (\citet{2000ApJ...544L.145H},\citet{2006AJ....132.2206B})
and where there are also asymmetries in the circumstellar debris disk,
which could be due to a much wider planet; such a wide planet might be
detectable with AO imaging, but is not yet detected - neither in \citet{2007AJ....133.2442J}
nor in our even deeper imaging observation shown in Fig. 1.

For any AO images with simple imaging (shift+add), or also when using
a semi-transparent coronagraph and/or a Laser Guide Star,
one can then also subtract the Point Spread Function (PSF) of the
bright host star after the shift+add procedure, in order to improve the dynamic
range, i.e. to improve the detection capability for very small separations (see Fig. 4).
For {\em PSF subtraction}, one can either use another similar nearby
star observed just before or after the target (as done e.g. in
the detection of $\beta$ Pic b, \citet{2009A&A...493L..21L}) or one can
measure the actual PSF of the host star in the shift+add image
and then subtract it.

Moreover one can obtain the very highest
angular resolutions at the diffraction limit by using sparse aperture interferometric masks in addition to AO. While very
good dynamic ranges can be achieved very close to stars, the size of the apertures in masking interferometry is limited by
the number of holes which are needed, in order to preserve non-redundancy, thus limiting the total reachable dynamic range.
Currently reached detection limits at VLT can
be found e.g.~in \citet{2011A&A...532A..72L}, beginning to reach the upper mass limits for planets given above.

In order to reduce present quasistatic PSF noise further one can use another technique called Angular Differential Imaging
(ADI) \citep{2006ApJ...641..556M}. Using this method a sequence of images is acquired with an altitude/azimuth telescope,
while the instrument field derotator is switched off, being the reason for the technique's alias name Pupil Tracking (PT).
This keeps the instrument and telescope optics aligned and allows the field of view to rotate with respect to the instrument.
For each image, a reference PSF is constructed from other appropriately selected images of the
same sequence and subtracted before all residual images are then rotated to align the field and are combined.\\
This technique was further improved by introducing an improved algorithm for PSF subtraction in combination with the ADI.
\citet{2007ApJ...660..770L} present this new algorithm called 'locally optimized combination of images' (LOCI).

While the ADI is inefficient at small angular separations, the simultaneous Spectral Differential Imaging (SDI) technique
offers a high efficiency at all angular separations. It consists in the simultaneous acquisition of images in adjacent narrow
spectral bands within a spectral range where the stellar and planetary spectra differ appreciably
\citep[see][and references therein]{2007ApJ...661.1208L}.

Moreover different kinds of phase masks are in use and are especially effective in combination with Adaptive Optics and
a coronagraph. Recently \citet{2010ApJ...722L..49Q} presented first scientific results using the Apodizing Phase Plate
coronagraph (APP) on VLT NACO to detect $\beta$ Pic b at 4 $\mu$m.

One can also detect planets as companions to normal stars with optical imaging
from a {\em space telescope} like the Hubbe Space Telescope, see e.g. \citet{2008Sci...322.1345K}
for the images of the planet Fomalhaut b. From outside the Earth atmosphere, there is
no atmospheric seeing, so that one can always reach the diffraction limit.

Previous reviews of AO imaging of planets were published in \citet{2008NewAR..52..117D},
\citet{2009ARA&A..47..253O}, and \citet{2010A&ARv..18..317A}.
Previous homogeneous mass determinations of planets and candidates imaged
directly were given in \citet{2008msah.conf..183N} and \citet{2009AIPC.1158..231S}.

\begin{figure}
\centering
\includegraphics[width=12cm]{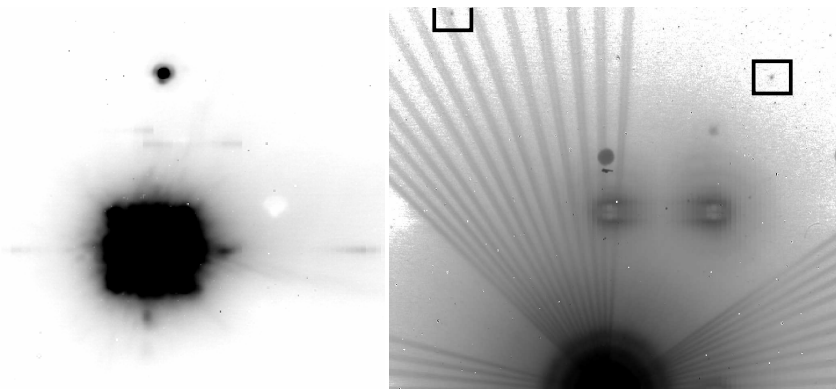}
\caption{Left: Our latest AO image of TWA 5 A (center) and B (2 arc sec
north of A) obtained with VLT/NACO on 2008 June 13 in the K band.
The mass of the companion is in the range of 17 to 45 M$_{\rm Jup}$
according to the \citet{1997ApJ...491..856B} model (Table 2).
Right: Our recent AO image of $\epsilon$ Eri obtained with VLT/NACO with
the star located behind a semi-transparent coronagraph. Due to the large
brightness of the star, reflection effects are also strong. Several very faint
objects are detected around $\epsilon$ Eri (see boxes); however, they are all non-moving
background objects as found after two epochs of observations; there is no
planet nor planet candidate detected, yet, nor any additional faint object
with only one epoch of imaging observation. Asymmetries in the debris
disk around $\epsilon$ Eri might be due to a wide planet.}
\end{figure}

\section{Proper motion confirmation of candidates}

Once a faint object is directly detected close to a star, one can consider it a
planet {\em candidate}, which needs to be confirmed. Two common tests
can be performed on such candidates: \\
(a) Common proper motion test: Both the star and the planet have
to show the same (or at least very similar) proper motion. The host star is normally
a relatively nearby star (up to a few hundred pc, otherwise the planet would be too
faint, i.e. not detectable), so that its proper motion is normally known.
If the faint object would be a background
star, it would be 1 to several kpc distant, so that its proper motion should be
negligible compared to the star. Hence, if both the star and the faint object show the
same proper motion, then the companion is not a non-moving background star,
but a co-moving companion. Given the orbital motion of the star and its companion,
depending on the inclination and eccentricity, one would of course expect that their
proper motions are not identical, but the differences (typically few milli arc sec
per year, mas/yr) are negligible compared to the typical proper motions.
Instead of (or best in addition to) common proper motion, it is also sufficient to show that
both objects (primary and companion candidate) show the same radial velocity,
and that the secular evolution of the radial velocity is consistent with orbital
motion and inconsistent with the background hypothesis. \\
(b) Spectrum: If the faint object next to the star would be a planet, its mass
and temperature should be much smaller than for the star. This can be shown by
a spectrum. Once a spectrum is available, one can determine the spectral type
and temperature of the companion. If those values are consistent with planetary
masses, then the faint object is most certainly a planet orbiting that star.
However, it could still be a very low-mass cool background object (very low-mass
L-type star or L- or T-type brown dwarf). In cases where the companion is too
faint and/or too close to the star, a spectrum might not be possible, yet, so
that one should try to detect the companion in different bands to measure color indices,
which can also yield (less precise) temperature or spectral type; then, however, one has
the problem to distinguish between a reddened background object and the
truely red (i.e. cool, e.g. planetary) companion. \\
The case of the ScoPMS 214 companion candidate (no. 1 or B) has shown that both tests
are neccessary for a convincing case: The young K2 T Tauri star ScoPMS 214, member
of the Scorpius T association, shares apparently common proper motion with a faint
object nearby (3 arc sec separation) over five years; however, a spectrum of this
companion candidate has shown that it is a foreground M dwarf \citep{2009ApJS..181...62M}.
Hence, the spectroscopic test is indeed necessary. Also, red colors alone
(even if together with common proper motion) is not convincing, because a
faint object near a star could just be reddened by extinction (background)
instead of being intrinsically red, i.e. cool.

It is not sufficient to show that a star and a faint object nearby show the same
proper motion (or proper motion consistent within 1 to 3 $\sigma$), one also has to show
that the data are inconsistent with the background hypothesis.
Common proper motion can be shown with two imaging detections with an epoch difference
large enough to significantly reject the background hypothesis, namely that the faint
object would be an unrelated non-moving background object. The epoch difference needed
depends on the astrometric precision of the detector(s) used and the proper motion
of the host star. We show an example in Fig. 2 and 3.
Spectra are usually taken with an infrared spectrograph with a large telescope and AO.

\begin{figure}
\centering
\includegraphics[width=10cm,angle=270]{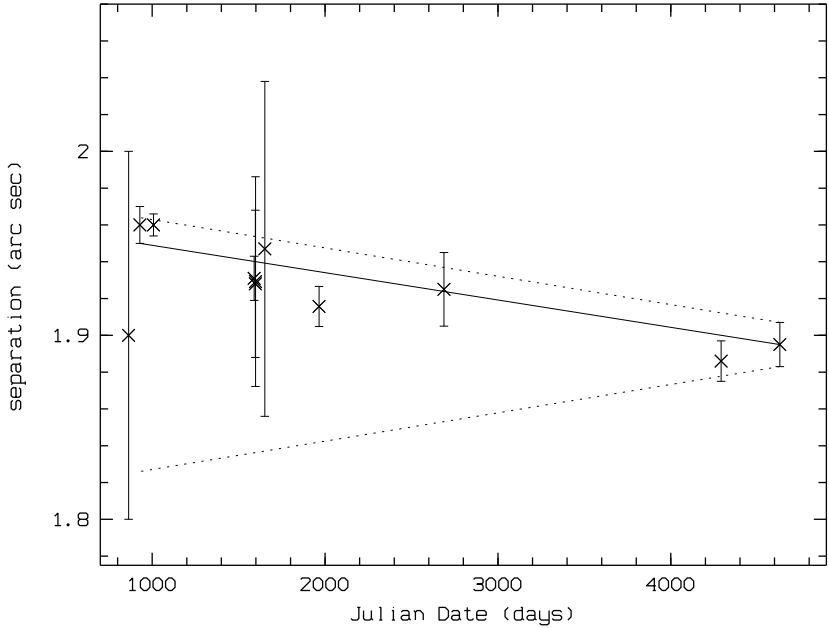}
\caption{Separation (in arc sec) versus observing epoch (JD - 2450000 in days)
between the host star TWA 5 A (actually the photo-center of close Aa+B pair)
and the sub-stellar companion TWA 5 B
using data from \citet{2010A&A...516A.112N}.
The dotted lines (starting from the 2008 data point opening to the
past) indicate maximum possible separation change due to orbital motion
for a circular edge-on orbit. The expectation for the background hypothesis
is not shown for clarity (and is rejected in Fig.~3).
All data points are fully consistent with common proper motion,
but not exactly identical proper motion (constant separation).
Instead, the data are fully consistent with orbital motion:
The separation decreases on average by $\sim 5.4$ mas per year,
as shown by the full line, which is the best fit.
The figure is adapted from \citet{2010A&A...516A.112N}.}
\end{figure}

\begin{figure}
\centering
\includegraphics[width=9cm,angle=270]{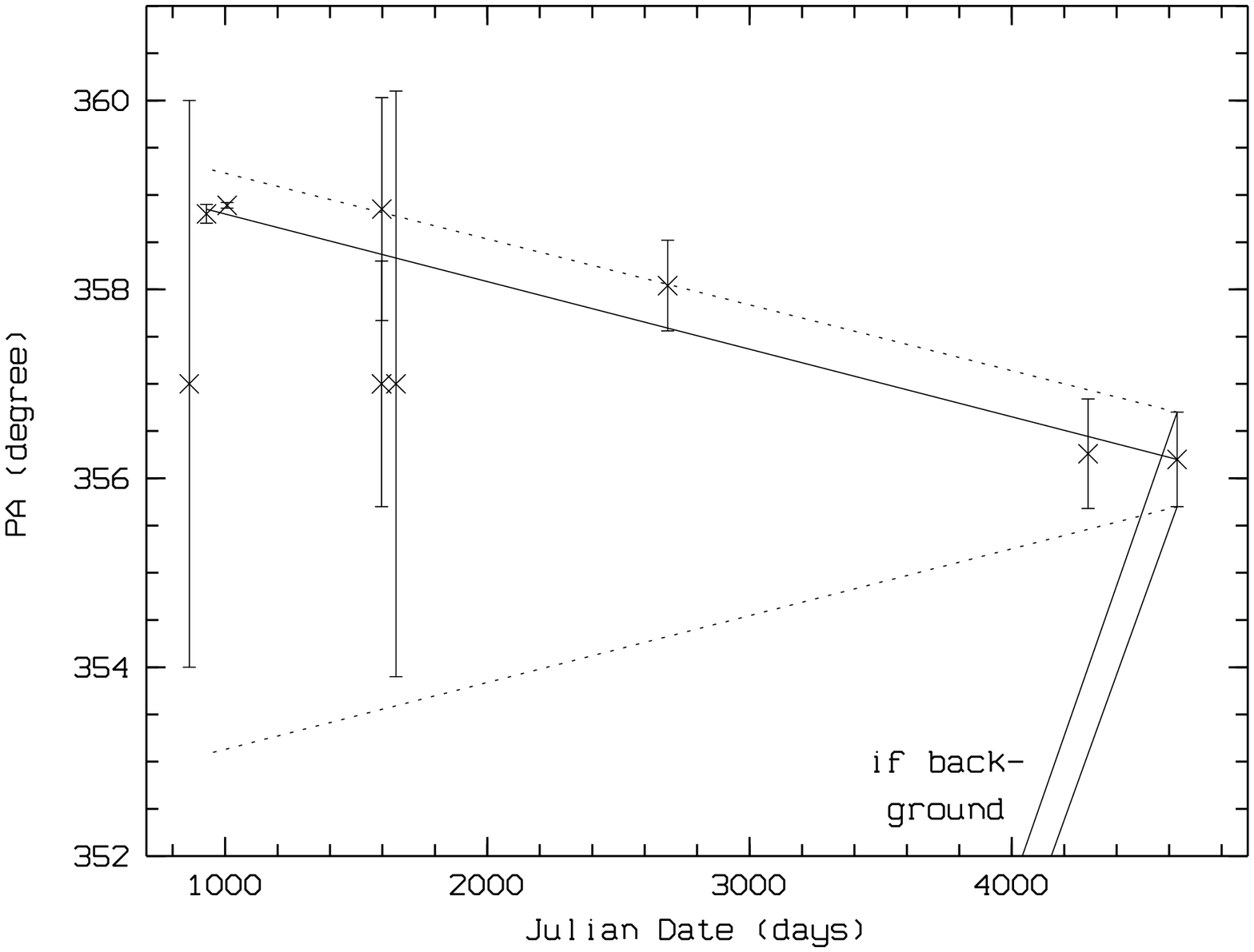}
\caption{Position angle PA (in degrees) versus observing epoch (JD - 2450000 in days)
for TWA 5 B with respect to TWA 5 A (actually the photo-center of close Aa+B pair)
using data from \citet{2010A&A...516A.112N}.
The dotted lines (starting from the 2008 data point opening to the past)
indicate maximum possible PA change due to orbital motion
for a circular pole-on orbit. The full lines with strong positive
slope in the lower right corner are for the background
hypothesis, if the bright central star (TWA 5 A) moved
according to its known proper motion, while the fainter
northern object (now known as B) would be a non-moving background object;
the data points are inconsistent with the background hypothesis
by many $\sigma$.
All data points are fully consistent with common proper motion,
but not exactly identical proper motion (constant PA).
Instead, the data are fully consistent with orbital motion:
The PA appears to decrease by $\sim 0.26^{\circ}$ per year,
as shown by the full line, which is the best fit.
The figure is adapted from \citet{2010A&A...516A.112N}.}
\end{figure}

Fig.~2 shows the change in separation between TWA 5 A and B with time,
Fig.~3 shows the position angle (PA) of TWA 5 B with respect to TWA 5 A.
In Fig.~3, the expectation for the background hypothesis is also plotted
and clearly rejected by many $\sigma$. All data points are consistent with
TWA 5 A and B being a common proper motion pair. Both the PA and the separation
values decrease since the first detection in 1999.
Such a (small) change in separation and/or PA can be interpreted as evidence
of orbital motion, which of course has to be expected.
One can conclude from these data that the orbit is eccentric and/or inclined.

A detection of a change in either separation or PA is actually
a detection of a difference in the proper motions of A and B.
It can be interpreted as evidence for slightly different proper motion.
Given that most directly imaged planets (or candidates) are
detected close to members of young associations (like TWA, Lupus, $\beta$ Pic
moving group etc.), it is therefore also possible that both the
host star A and the faint object nearby (B or b) are both independent
members of that association, not orbiting each other. Such an
association is partly defined by the way that all or most members show
a similar proper motion. In such cases, it might be necessary to
show not only common proper motion (i.e. similar proper motion within
the errors) or slightly different proper motion (consistent with
orbital motion), but also curvature in the orbital motion.
Such curvature would be due to acceleration (or deceleration)
in case of a non-circular orbit. It could also be due to apparent
acceleration (or deceleration) in the 2-dimensional apparent
orbit on the plane of the sky for an orbit that is inclined towards
the plane of the sky. Curvature can also be detected if the faint object
is not anymore a bound companion, but is currently been ejected,
i.e. on a hyperbolic orbit. Hence, to convincingly prove that a faint
object is a bound companion, one has to show curvature that is
not consistent with a hyperbolic orbit.

For all of the directly imaged planets and candidates listed below,
common proper motion between the candidate and the host star has been shown.
For only few of them, evidence for orbital motion is shown, e.g.
DH Tau \citep{2005ApJ...620..984I}
GQ Lup \citep{2008A&A...484..281N}, Fomalhaut \citep{2008Sci...322.1345K},
HR 8799 bcde \citep{2008Sci...322.1348M,2010Natur.468.1080M},
TWA 5 \citep{2010A&A...516A.112N}, PZ Tel \citep{2010A&A...523L...1M},
$\beta$ Pic \citep{2010Sci...329...57L}, and HR 7329 \citep{2011MNRAS.tmp.1135N}.
For only two of them, curvature in the orbital motion was detected,
namely in PZ Tel \citep{2010A&A...523L...1M} and TWA 5 \citep{2010A&A...516A.112N}.

\section{Data base: Planets and candidates imaged directly}

Searching the literature, we found 25 stars with directly imaged planets and candidates.
We add 2M1207, a planet candidate imaged directly, whose primary would be a brown dwarf.
In two cases, there is more than one planet (or candidate) detected directly to
orbit the star: 4 planets around HR 8799 and two candidates in the GJ 417 system, see below.
Most planets and candidates orbit single stars, but there are some as members of
hierachical systems with three or more objects (one planet candidate plus two or more
stars, such as TWA 5 or Ross 458).
We gathered photometric and spectral information for all these objects,
to derived their luminosities in a homogeneous way,
taking a bolometric correction into account (Table 1).
According to the mass estimate in Table 2, all of them
can have a mass below 25~M$_{\rm Jup}$, so that they are considered as planets.

The masses of such companions can be determined in
different ways, the first two of which are usually used:
\begin{itemize}
\item Given the direct detection of the companion, its brightness is measured. If the companionship
to the star is shown, e.g. by common proper motion, then one can assume that the companion
has the same distance and age as its host star. If either a spectrum or color index is also
observed, one can estimate the temperature of the companion, so that the bolometric correction
can be determined; if neither color nor spectrum is available, one can often roughly estimate
the temperature from the brightness difference, assuming companionship. From brightness,
bolometric correction, and distance, one can estimate the luminosity. Using theoretical evolutionary
models, one can then estimate the mass from luminosity, temperature, and age. However, those models
are uncertain due to unknown initial conditions and assumptions. In particular for the youngest
systems, below 10 Myr, the values from the models are most uncertain. Masses derived in this
way are listed in Table 2.
\item If a good S/N spectrum with sufficient resolution is obtained, one can also measure
the effective temperature and surface gravity of the companion. Then, from temperature and
luminosity, one can estimate the companion radius. Then, from radius and gravity,
one can estimate the companion mass. This technique is independent of the uncertain
models, but needs both distance and gravity with good precision.
Since gravities (and sometimes also distances) cannot be measured
precisely, yet, the masses derived in this way typically have a very large possible range.
\item In the case of the directly imaged planet around the star Fomalhaut,
an upper mass limit of $\sim 3$~M$_{\rm Jup}$
for the companion could be determined by the fact that a dust debris ring is located
just a few AU outside the companion orbit \citep{2008Sci...322.1345K}.
In other planet candidates also orbiting host stars with debris disks,
such an upper mass limit estimate should also be possible,
e.g. in HR 8799 (\citet{2008Sci...322.1348M},\citet{2009A&A...503..247R}),
$\beta$ Pic \citep[see][]{2007A&A...466..389F},
PZ Tel (\citet{2010ApJ...720L..82B},\citet{2010A&A...523L...1M}), and HR 7329 \citep{2011MNRAS.tmp.1135N}.
\item If there are several planets or candidates imaged around the same star,
then one can also try to determine masses or limits by stability arguments, see e.g. HR 8799
(\citet{2008Sci...322.1348M},\citet{2009A&A...503..247R}).
\item If there are other sub-stellar objects with very similar values regarding
temperature, luminosity, and age, for which there is also a direct mass estimate,
e.g. directly obtained in an eclipsing double-lined binary such as 2M0535 \citep{2006Natur.440..311S}
or in a visually resolved system with full orbit determination such as HD 130948 BC \citep{2009ApJ...692..729D},
one can conclude that the sub-stellar companion in question also has a similar mass.
If a sub-stellar companion has temperature and luminosity smaller than another
sub-stellar object with a direct mass estimate, but the same age, then the sub-stellar
companion in question should have a smaller mass.
For HD 130948 BC, there is only an estimate for the total mass being $114 \pm 3$~M$_{\rm Jup}$
\citep{2009ApJ...692..729D}, i.e. somewhat too large for comparison to planet candidates.
The object 2M0535 is an eclipsing double-lined spectroscopic binary comprising of
two brown dwarfs, member of the Orion star forming region, hence not older than a
few Myr, maybe below 1 Myr. For a double-lined spectroscopic binary, one can
determine brigtness, temperatures, luminosities, and lower mass limits $m \cdot \sin i$
for both objects individually.
The orbital inclination $i$ can then be obtained from the eclipse light curve.
Hence, both masses are determined dynamically without model assumptions,
the masses are $60 \pm 5$~M$_{\rm Jup}$ for A and $38 \pm 3$~M$_{\rm Jup}$ for B \citep{2007ApJ...664.1154S}.
Given that several of the sub-stellar companions discussed here have a very
similar age, we can compare them with 2M0535 A and B. If all parameters are similar,
than the masses should also be similar. If a companion has lower values (at a similar age),
i.e. being cooler and fainter, then it will be lower in mass.
See Table 1 and 2 for the values and the comparison.
Such a comparison should also be done with great care, because also other properties
like magnetic field strength, spots on the surface, and chemical composition (metallicity)
affect the analysis.
\item If one could determine a full orbit of two objects around each other,
one can then estimate the masses of both the host star and the companion using
Kepler's 3rd law as generalized by Newton. However, since all planets and planet
candidates imaged directly so far have large separations ($\ge 8.5$ AU) from their host star
(otherwise, they would not have been detected directly), the orbital periods
are typically tens to hundreds of years, so that full orbits are not yet observed.
\end{itemize}

\section{Comments on individual objects}

Here, we list data, arguments, and problems related to the classification of the companions
as planets. We include those sub-stellar companions, where the common proper motion
with a {\em stellar} primary host star has been shown with at least $3~\sigma$ and where
the possibly planetary nature, i.e. very low mass and/or cool temperature,
has been shown by a spectrum - or at least a very red color with known (small) extinction,
in order to exclude reddened background objects.
We also include the brown dwarf 2M1207 with its fainter and lower-mass sub-stellar companion,
even though the primary object is not a star, but we do not list other brown dwarfs with
possibly planetary mass companions. We include only those systems in our list for
new and homogeneous mass determination, where the age is considered
to be possibly below $\sim 500$ Myr, otherwise age and sub-stellar companion mass is probably too large;
however, we do include those older systems, where the mass of the sub-stellar companion has already been
published and estimated to be near or below $\sim 25$~M$_{\rm Jup}$, e.g. WD 0806-661 \citep{2011ApJ...730L...9L}
and Wolf 940 B \citep{2009MNRAS.395.1237B}.
We also exclude those systems, however, where the age is completely
unconstrained, e.g. HD 3651 \citep{2006MNRAS.373L..31M}, GJ 758 \citep{2009ApJ...707L.123T},
and several others listed, e.g., in \citet{2010AJ....139..176F}.
This compilation is the 3rd version (after \citet{2008msah.conf..183N} and \citet{2009AIPC.1158..231S}) and
we do plan to renew and enlarge the catalog later; then, we will consider to also include
possibly planetary mass companions imaged directly around sub-stellar primaries and
around old stars or stars with unconstrained age.

We list the objects in the chronologic order of the publications of the common proper motion
confirmations; if the significant confirmation were published later than the discovery,
we list the object at the later date.

\begin{table}
\centering
\begin{tabular}{lccccl}
\hline
Object    & Luminosity                     & Magnitude         & Temp.             & Age   & Comments \\
name      & $\log (L_{\rm bol}/L_{\odot})$ & M$_{\rm K}$ [mag] & T$_{\rm eff}$ [K] & [Myr] & \\ \hline
\multicolumn{6}{c}{Reference object (eSB2 brown dwarf - brown dwarf binary 2M0535):} \\
\hline
2M0535 A  & $-1.65 \pm 0.07$ & $5.29 \pm 0.16$ & $2715 \pm 100$ & 1 (0-3) & \citet{2007ApJ...664.1154S} \\
       B  & $-1.83 \pm 0.07$ & $5.29 \pm 0.16$ & $2820 \pm 105$ & 1 (0-3) & \citet{2007ApJ...664.1154S} \\
\hline
\multicolumn{6}{c}{Directly detected planet candidates:} \\
\hline
GG Tau Bb     & $-1.84 \pm 0.32$     & $ 6.28 \pm 0.79$ & $2880 \pm 150$ & 0.1-2  & (1), (2)         \\
TWA 5 B       & $-2.62 \pm 0.30$     & $ 8.18 \pm 0.28$ & $2800 \pm 450$ & 5-12   &                  \\
GJ 417 B \& C & $-4.14 \pm 0.06$   & $11.74 \pm 0.05$ & $1600 \pm 300$ & 80-300 & each object (1), (3) \\
GSC 08047 B/b & $-3.58 \pm 0.28$     & $10.75 \pm 0.60$ & $2225 \pm 325$ & 25-40  &                  \\
DH Tau B/b    & $-2.81 \pm 0.32$     & $ 8.46 \pm 0.78$ & $2750 \pm 50$  & 0.1-10 &                  \\
GQ Lup b      & $-2.25 \pm 0.24$     & $ 7.37 \pm 0.78$ & $2650 \pm 100$ & 0.1-2  &                  \\
2M1207 b      & $-4.74 \pm 0.06$     & $13.33 \pm 0.12$ & $1590 \pm 280$ & 5-12   &                  \\
AB Pic B/b      & $-3.73 \pm 0.09$     & $10.82 \pm 0.11$ & $2000^{+100}_{-300}$ & 25-40 & (1)         \\
LP 261-75 B/b   & $-3.87 \pm 0.54$     & $11.18 \pm 1.34$ & $1500 \pm 150$ & 100-200 & (1)             \\
HD 203030 B/b   & $-4.64 \pm 0.07$     & $13.14 \pm 0.12$ & $1440 \pm 350$ & 130-400 & T$_{\rm eff}$ error (3) \\
HN Peg B/b      & $-4.93 \pm 0.16$     & $14.37 \pm 0.25$ & $1450 \pm 300$ & 200-300 & (1), (3)        \\
CT Cha b        & $-2.68 \pm 0.21$     & $ 8.86 \pm 0.50$ & $2600 \pm 250$ & 0.1-4   & (7)             \\
Fomalhaut b     & $\le -6.5$           & M$_{\rm H} \ge 23.5$ &            & 100-300 & no colors/spectra \\
HR 8799 b       & $-5.1 \pm 0.1$       & $14.05 \pm 0.08$ & $1300 \pm 400$ & 20-1100 &                 \\
\ \ \ \ \ \ \ \ \ \ \ \ \ \ \ \
            c   & $-4.7 \pm 0.1$       & $13.13 \pm 0.08$ & $\sim 1100$    & 20-1100 &                 \\
\ \ \ \ \ \ \ \ \ \ \ \ \ \ \ \
            d   & $-4.7 \pm 0.1$       & $13.11 \pm 0.12$ &                & 20-1100 &                 \\
\ \ \ \ \ \ \ \ \ \ \ \ \ \ \ \
            e   & $-4.7 \pm 0.2$       & $12.93 \pm 0.22$ &                & 20-1100 &                 \\
Wolf 940 B/b    & $-6.07 \pm 0.04$     & $18.36 \pm 0.16$ & $ 600 \pm 100$ & 3.5-6 Gyr &               \\
G 196-3   B/b   & $-3.8^{+0.2}_{-0.3}$ & $11.17 \pm 0.62$ & $1870 \pm 100$ & 20-600  &                 \\
$\beta$ Pic b   & $-3.903^{+0.074}_{-0.402}$ & $11.20 \pm 0.11$ & $1700 \pm 300$ & 8-20 & no spectra, (1), (4) \\
RXJ1609 B/b     & $-3.55 \pm 0.20$     & $10.36 \pm 0.35$ & $1800^{+200}_{-100}$ & 5 (1-10) &          \\
PZ Tel B/b      & $-2.58 \pm 0.08$     & $ 8.14 \pm 0.15$ & $2600 \pm 100$ & 8-20    &  (5), (1)       \\
Ross 458 C      & $-5.62 \pm 0.03$     & $16.11 \pm 0.05$ & $ 650 \pm  25$ & 150-800 &                 \\
GSC 06214 B/b   & $-3.09 \pm 0.12$     & $ 9.17 \pm 0.23$ & $2050 \pm 450$ & 5 (1-10)& (1), (3)        \\
CD-35 2722 B/b  & $-3.59 \pm 0.09$     & $10.37 \pm 0.16$ & $1800 \pm 100$ & 50-150  & (1)             \\
HIP 78530 B/b   & $-2.55 \pm 0.13$     & $ 8.19 \pm 0.18$ & $2800 \pm 200$ & 5 (1-10)&                 \\
WD 0806-661 B/b &                      & M$_{\rm J} \ge 21.7$ & $300 $     & 1.5-2.7 Gyr& (6)          \\
SR 12 C         & $-2.87 \pm 0.20$     & $ 9.09 \pm 0.44$ & $2400^{+155}_{-100}$ & 0.3-10 &            \\
HR 7329 B/b     & $-2.63 \pm 0.09$     & $ 8.21 \pm 0.12$ & $2650 \pm 150$ & 8-20   &                  \\
%HD 49197 b    & $-3.86 \pm 0.09$     & $11.03 \pm 0.15$ & $1850 \pm 500$ & 260-790 & (1), (3)        \\
%DENIS-P J1347-7610 b
%              & $-3.45 \pm 0.25$     & $10.17 \pm 0.61$ & $2300 \pm 150$ & 200-1400 & (1), (3)       \\
%              &                      &                  & $1000 \pm 100$ & 5-12   & Barman 11        \\
%Oph 1622 a   & $-2.61 \pm 0.23$ & $7.96 \pm 0.56$  &                & 5 (1-20)& Allers 2006, [2] \\
%Oph 1622 b   & $-2.79 \pm 0.23$ & $8.42 \pm 0.56$  &                & 5 (1-20)& [8], [15]. [18]. [24] \\
%SCR 1845 b   & \multicolumn{2}{c}{M$_{\rm H} = 15.30\,+0.31\,-0.26$}& (850)   & 100-10$^{4}$& Biller 2006 \\
%CHXR 73 b    & $-2.62 \pm 0.21$ & $8.00 \pm 0.52$  & $2600 \pm 450$ & 0.1-4   & Luhman 2006, [33] \\
%USco 108 b    & $-3.14 \pm 0.14$ & $9.30 \pm 0.34$  & (2350)         & 5 (1-10)& B{\'e}jar 2008 \\
%FU Tau b     & $-2.40 \pm 0.09$ & $7.44 \pm 0.20$  &                & 0.1-4   & Luhman 2009 \\
\hline
\end{tabular}
\caption{Observed properties of the directly imaged planet candidates. References given in Sect.~5 (1) luminosity from
spectral type (BC from Golimowski et al. (2004)), K magnitude and distance (2) temperature from spectral type using
Luhman et al. (2003) (3) temperature from spectral type using Golimowski et al. (2004) (4) temperature from Ks - L' color
(5) temperature from JHK colors (6) detected at 4.5 $\mu$m (7) Only for CT Cha b extinction was taken into account as given 
in Schmidt et al. (2008)}
\end{table}

{\bf GG Tau Bb}: The T Tauri star GG Tau in the Taurus T association (hence 0.1 to 2 Myr young
at $\sim 140 $ pc distance) is a quadruple system with two close pairs
GG Tau Aa+Ab (separation $\sim 0.25 ^{\prime \prime} \simeq 35$ AU,
the fainter component in this northern binary is sometimes called GG Tau a)
and GG Tau Ba+Bb ($\sim 1.48 ^{\prime \prime} \simeq$ 207 AU),
the separation between A and B is $10 ^{\prime \prime}$.
The system has been studied in detail by \citet{1999ApJ...520..811W}
using the HST as well as HIRES and LRIS at Keck.
The object of interest here is GG Tau Bb, also called GG Tau/c.
\citet{1999ApJ...520..811W} determine a spectral type of M7, zero extinction, Lithium
absorption, and H$\alpha$ emission (hence, young, not a reddened background object).
According to \citet{1998A&A...337..403B} models, at the age and distance of the star GG Tau Ba,
the sub-stellar object GG Tab Bb has a mass of 40 to 60~M$_{\rm Jup}$,
or only 20 to 50~M$_{\rm Jup}$ according to \citet{1994ApJS...90..467D,1997MmSAI..68..807D}.
Also \citet{2001A&A...376..982W} give 20 to 40~M$_{\rm Jup}$ for GG Tau Bb
using the \citet{1998ASPC..134..442D} and \citet{1998A&A...337..403B} models.
We compared its properties with the \citet{1997ApJ...491..856B} models,
where the mass can be as low as $\sim 23$~M$_{\rm Jup}$,
so that we include the object in this study.
Kinematic confirmation was done by \citet{1999ApJ...520..811W}, who could determine
the radial velocities of both GG Tau Ba and Bb to be $16.8 \pm 0.7$ km/s
and $17.1 \pm 1.0$ km/s, i.e. consistent with each other and also with GG Tau A.
The pair GG Tau A+B is also a common proper motion pair according to
the NOMAD \citep{2005yCat.1297....0Z} and WDS catalogs \citep{2001AJ....122.3466M}.
If accepted as planet, it would be the first planet imaged directly and confirmed by
both common proper motion and spectroscopy.
Orbital motion or curvature in orbital motion of GG Tau Bb around Ba were not yet reported.
% 1999

{\bf TWA 5}: The first direct imaging detection of the companion $1.960 \pm 0.006 ^{\prime \prime}$
($86.2 \pm 4.0$ AU) off TWA 5 (5 to 12 Myr as member of the TW Hya association, $44 \pm 4$ pc, M1.5)
was done with NASA/IRTF and Keck/LRIS by \citet{1999ApJ...512L..63W} and with HST/Nicmos
and Keck/NIRC by \citet{1999ApJ...512L..69L}.
The common proper motion and spectral (M8.5-9) confirmation was given in \citet{2000A&A...360L..39N},
who derived a mass of 15-40~M$_{\rm Jup}$ from formation models at the age and distance of the star.
The mass lies anywhere between 4 and 145~M$_{\rm Jup}$, if calculated from temperature ($2800 \pm 100$ K),
luminosity ($\log(L_{bol}/L_{\odot}) = -2.62 \pm 0.30$ at $44 \pm 4$ pc),
and gravity ($\log g = 4.0 \pm 0.5$ cgs), as obtained by comparison of a
VLT/Sinfoni K-band spectrum with Drift-Phoenix model atmospheres \citep{2009AIPC.1094..844N}.
The temperature error given in \citet{2009AIPC.1094..844N} may be underestimated, because it is
only from the K-band, so that we use a conservative, larger error here in Table 2.
In our Table 2 below, we give 17 to 50~M$_{\rm Jup}$ as possible mass range.
For any of those mass ranges, it might well be below 25~M$_{\rm Jup}$,
hence it is also a planet candidates imaged directly
(called here TWA 5 B, but not b, in order not to confuse with TWA 5 Ab).
Our latest image of TWA 5 A+B is shown in Fig. 1.
Orbital motion of B around A was shown in \citet{2010A&A...516A.112N};
the host star A is actually a very close 55 mas binary star.
The data for separation and position angle plotted in Fig. 2 and 3 here
are corrected for the binarity of the host star, i.e. are the values
between the companion B and the photocenter of Aa+Ab using the orbit
of Aa+b from \citet{2007AJ....133.2008K}.
What was plotted as orbital motion in figures 1 and 2 in \citet{2010A&A...516A.112N},
actually is a small difference in proper motions of TWA 5 A and B,
so small, that it is consistent with the expected orbital motion;
figures 1 and 2 in \citet{2010A&A...516A.112N} only show a linear fit.
However, true orbital motion cannot be linear, but always shows some curvature.
In figure 3 in \citet{2010A&A...516A.112N}, it is shown from geometric fits
that an eccentric orbit of B around A is most likely, hence curvature is
detected (with low significance).
Curvature is not yet a final proof for being bound, because
it would also be expected for an hyperblic orbit.
The orbital period is $\sim 950$ yr (for a circular orbit) or $\sim 1200$ yr for an
eccentric orbit with the best fit eccentricity $e=0.45$ \citep{2010A&A...516A.112N}.
% 2000

{\bf GJ 417}: Common proper motion between the binary star GJ 417
(or Gl 417, CCDM J11126+3549AB, WDS J11125+3549AB, called primary A in \citet{2003AJ....126.1526B})
and the companion 2MASS~J1112256+354813 was noticed by \citet{2003AJ....126.1526B}.
The primary, GJ 417, is comprised by two stars both with spectral type G0 (Simbad).
This primary GJ 417 A (or GJ 417 Aa+b) has common proper motion with
2MASS~J1112256+354813 at a wide separation of $\sim 90 ^{\prime \prime}$
(or $\sim 1953$ AU at $\sim 21.7$ pc, \citet{2003AJ....126.1526B}).
The secondary, 2MASS~J1112256+354813, is a close binary itself
(called B and C in \citet{2003AJ....126.1526B})
with $\sim 0.0700 ^{\prime \prime}$ separation with almost
equal magnitudes and a combined spectral type of L4.5 \citep{2003AJ....126.1526B}.
At the distance of GJ 417, this separation corresponds to only $\sim 1.5$ AU
projected separation and a very short period of few years.
The age of the system is given to be only 80 to 300 Myr \citep{2010AJ....139..176F},
so that the mass of B and C can be near the planetary mass regime \citep{2010AJ....139..176F}.
%Sept 2003

{\bf GSC 08047} (GSC~08047-00232): The first direct imaging detections of this
companion $3.238 \pm 0.022 ^{\prime \prime}$ ($219 \pm 59$ AU)
off GSC 08047 (K2, 50-85 pc, 25 to 40 Myr as member of the TucHor Association)
was shown in \citet{2003AN....324..535N} using simple IR imaging with NTT/SofI
and IR speckle imaging with NTT/Sharp as well as in \citet{2003A&A...404..157C} with AO
imaging using NTT/Adonis. \citet{2004A&A...420..647N} could show common proper motion,
while both \citet{2004A&A...420..647N} and \citet{2005A&A...430.1027C} presented spectra
(M6-9.5). Based on formation models, \citet{2004A&A...420..647N}
derived the mass of the companion to be 7-50~M$_{\rm Jup}$ at the age and distance of the star.
% A&A 420 in 2004

{\bf DH Tau}: Direct imaging AO detection with Subaru/CIAO of the companion
$2.351 \pm 0.001 ^{\prime \prime}$ ($\sim 329$ AU) off DH Tau (0.1 to 10 Myr and at $\sim 140$ pc
as member of the Taurus T association, M0.5)
were given in \citet{2005ApJ...620..984I}, who also could show common proper motion and
a high-resolution spectrum with Subaru/CISCO giving temperature and gravity
yielding a mass of 30-50~M$_{\rm Jup}$.
A small difference observed in the position angle is consistent with orbital motion \citep{2005ApJ...620..984I}.
% Feb 2005

{\bf GQ Lup}: Direct imaging AO detection with VLT/NACO of the companion
$0.7347 \pm 0.0031 ^{\prime \prime}$ ($\sim 100$ AU) off GQ Lup
(0.1 to 2 Myr and at $\sim 140$ pc as member of the Lupus-I T association, K7)
together with common proper motion, spectral classification (late-M to early-L),
and a mass estimate of 1-42~M$_{\rm Jup}$ \citep{2005A&A...435L..13N}
were confirmed by \citet{2006A&A...453..609J} giving a mass of 3-20~M$_{\rm Jup}$,
\citet{2007ApJ...654L.151M} deriving 10-20~M$_{\rm Jup}$, and
\citet{2007ApJ...656..505M} listing 10-40~M$_{\rm Jup}$, all from photometry and temperature
of the companion, age, and distance of the star, together with formation models.
\citet{2007A&A...463..309S} obtained higher-resolution VLT/Sinfoni spectra to derive the
gravity of the companion and determined the mass model-independant to be 4-155~M$_{\rm Jup}$.
Evidence for a few mas/yr orbital motion \citep{2008A&A...484..281N} does not yet show curvature.
Hence, the companion to GQ Lup can be a massive planet or a low-mass brown dwarf.
% apr/May 2005

{\bf 2M1207} (2MASSWJ1207334-393254): While the first direct imaging AO detection with VLT/NACO
of this companion $0.7737 \pm 0.0022 ^{\prime \prime}$ ($40.5 \pm 1$ AU)
off the brown dwarf 2M1207 A
(5 to 12 Myr as member of the TW Hya Association, $52.4 \pm 1.1$ pc, M8 brown dwarf)
was published in \citet{2004A&A...425L..29C},
the proper motion confirmation was given in \citet{2005A&A...438L..25C}.
\citet{2005AN....326..629M} and \citet{2007ApJ...660.1492C} noticed that the binding energy between 2M1207 and
its companion may not be sufficient for being bound or for staying bound for a long time:
The total mass is too low for the large separation.
Orbital motion or curvature in the orbital motion were not yet shown.
The companion also appears to be too faint given its L5-L9.5 spectral type and 5 to 12 Myr age
\citep{2007ApJ...657.1064M}.
% May 2005

{\bf AB Pic}: The first direct imaging AO detection with VLT/NACO of this
companion $5.460 \pm 0.014 ^{\prime \prime}$ ($251.7 \pm 8.9$ AU)
off AB Pic ($46.1 \pm 1.5$ pc, K1, 25 to 40 Myr as member of the TucHor Association)
together with spectrum (L0-2), proper motion
confirmation, and a mass estimate based on formation models to be 13-14~M$_{\rm Jup}$ were published
in \citet{2005A&A...438L..29C}. \citet{2010A&A...512A..52B} obtained temperature and gravity with VLT/Sinfoni
spectra and estimated the mass to be 1 to 45~M$_{\rm Jup}$.
% may/June 2005

{\bf LP 261-75}: \citet{2006PASP..118..671R} noticed this common proper motion pair:
The companion 2MASSW J09510549+3558021 (or LP 261-75 B) has a
separation of $\sim 12^{\prime \prime}$ ($\sim 744$ AU)
off LP 261-75 ($62 \pm 38$ pc, M4.5, 100 to 200 Myr due to its activity).
\citet{2006PASP..118..671R} presented a spectrum of the companion (L6),
showed proper motion confirmation,
and estimated the mass based on formation models to be 15-30~M$_{\rm Jup}$.
% May 2006

{\bf HD 203030}: The first direct imaging AO detection with Palomar of this
companion $11.923 \pm 0.021 ^{\prime \prime}$ ($503 \pm 15$ AU)
off HD 203030 ($42.2 \pm 1.2$ pc, G8, 130 to 400 Myr due to its activity)
together with spectrum (L7-8), proper motion confirmation,
and a mass estimate based on formation models to be 12-31~M$_{\rm Jup}$
were published in \citet{2006ApJ...651.1166M}.
% Nov 2006

{\bf HN Peg}: First direct imaging detection with 2MASS and Spitzer/IRAC
of this companion $43.2 \pm 0.4 ^{\prime \prime}$ ($773 \pm 13$ AU) off
HN Peg ($17.89 \pm 0.14$ pc, G0, 200 to 300 Myr, if a member of the Her-Lyr group)
together with common proper motion confirmation were shown in \citet{2007ApJ...654..570L}.
Given the NASA/IRTF/SpecX spectral detection of methane, the companion can be classified T$2.5 \pm 0.5$,
so that it has 12-30~M$_{\rm Jup}$ at the age and distance of the star.
The large projected separation of $\sim 773$ AU may favor the brown dwarf interpretation.

%{\bf UScoCTIO 108}: The first direct imaging detection with 2MASS and WHT/AUX
%of this companion $ \pm ^{\prime \prime}$ ($ \pm $ AU) off
%UScoCTIO 108 (ps age ST)
%together with common proper motion confirmation and spectroscopy (M7-9.5)
%were published by Bejar et al. (2008), who gave a mass of 6-16~M$_{\rm Jup}$
%based on formation models for age and distance of the star.
%Orbital motion or curvature in the orbital motion were not yet shown.

{\bf CT Cha}: First direct imaging AO detection with VLT/NACO of this
companion $2.670 \pm 0.036 ^{\prime \prime}$ ($441 \pm 87$ AU) off
CT Cha ($165 \pm 30$ pc and 0.1 to 4 Myr as member of the Cha I T association, K7)
together with VLT/Sinfoni JHK spectra (M8-L0),
common proper motion confirmation, and a mass estimate based on formation models to be
11-23~M$_{\rm Jup}$ at the age and distance of the star
were published in \citet{2008A&A...491..311S}, so that this companion can also
be a high-mass planet or low-mass brown dwarf.

{\bf Fomalhaut b}: Direct imaging detection in the red optical with the Hubble Space
Telescope (HST) of this companion
$12.7 ^{\prime \prime}$ ($\sim 100$ AU) off
Fomalhaut ($7.704 \pm 0.028$ pc, A4, 100 to 300 Myr old)
together with common proper motion confirmation were published in \citet{2008Sci...322.1345K}.
They also estimated the mass of the companion to be below $\sim 3$~M$_{\rm Jup}$ due to its
location close to the dusty debris disk seen in reflected optical light.
\citet{2008Sci...322.1345K} also obtained a few imaging photometric points and upper limits;
those data points were not consistent with the expected spectrum of a low-mass cool planet,
so that they could not exclude that the emission is reflected light from the small cloud-let.
A spectrum or an IR detection of the companion could not yet be obtained.
After two imaging epochs, the slightly different positions of the companion
with respect to the star are consistent with orbital motion \citep{2008Sci...322.1345K},
but curvature in the orbital motion was not yet shown.

{\bf HR 8799}: Direct imaging AO detection with Keck/NIRC2 and Gemini North/NIRI of the companions b, c, and d
was shown by \citet{2008Sci...322.1348M} together with common proper motion confirmation
for all three companions, while orbital motion could be shown for companions b and c. The orbital motion confirmation
of the third candidate, d, was later given with higher significance by \citet{2009ApJ...705L.204M}. A fourth candidate,
HR 8799 e, was later detected by \citet{2010Natur.468.1080M},
whose orbital (and common proper motion, however no significance is explicitely given for this) with the star was shown
within this publication.
The companions b, c, d, and e have separations of $1.713 \pm 0.006 ^{\prime \prime}$,
$0.952 \pm 0.011 ^{\prime \prime}$, $0.613 \pm 0.026 ^{\prime \prime}$, and $0.362 \pm 0.033 ^{\prime \prime}$,
respectively, which correspond to 14 to 67 AU at the distance of the star being $39.4 \pm 1.0$ pc,
similar as the solar system dimension. Spectra were taken for HR 8799 b by \citet{2010ApJ...723..850B} and
\citet{2011ApJ...733...65B} and for HR 8799 c by \citet{2010ApJ...710L..35J},
showing tempartures of 1300 -- 1700 K and 1100 $\pm$ 100 K for HR 8799 b and $\sim$ 1100 K for HR 8799 c, respectively.
The age of the star is somewhat uncertain: Given its bright debris disk, it might be as young as
20 Myr, then the companions are certainly below 13~M$_{\rm Jup}$; astroseismology, however,
seem to indicate that the star can be as old as $\sim 1.1$ Gyr, then the companions would be brown dwarfs.

{\bf Wolf 940}: The first imaging detection in UKIDSS, later also detected with Keck AO and
laser guide star, of this companion $32^{\prime \prime}$ ($\sim 400$ AU) off
Wolf 490 ($12.53 \pm 0.71$ pc, 3.5-6 Gyr due to activity, M4)
was presented by \citet{2009MNRAS.395.1237B} together with common proper motion and spectra (T8.5).
Comparison of their spectra with BT Settl models yields 500 to 700 K,
while the temperature and gravity estimates also given in \citet{2009MNRAS.395.1237B}, $570 \pm 25$ K,
are obtained using the radius from cooling models for the age range from the stellar activity,
hence possibly less reliable. A mass was also obtained from models for the given age range,
namely 20-32~M$_{\rm Jup}$ \citep{2009MNRAS.395.1237B}.
% 22 Apr 2009

{\bf G 196-3}: The first direct imaging detection of the companion
$\sim 16.2^{\prime \prime}$ (243-437 AU)
off G186-3 (host star: M2.5V, 20 to 600 Myr, 15-27 pc)
was published by \citet{1998Sci...282.1309R} using the 1m NOT/ALFOSC and HiRAC instruments
in the red optical and IR without AO, together with common proper motion ($2~\sigma$ only) and
spectroscopic (L3) confirmation. The mass
was estimated to be 15-40~M$_{\rm Jup}$ or 12-25~M$_{\rm Jup}$ from
cooling models for 20 to 600 Myr \citep{1998Sci...282.1309R} or 20 to 300 Myr \citep{2010ApJ...715.1408Z}.
if the $2~\sigma$ common proper motion is
accepted as confirmation, then is would actually be the first imaged planet (candidate) confirmed as
co-moving companion by proper motion \citep{1998Sci...282.1309R}; higher significance for common
proper motion was presented by \citet{2010ApJ...715.1408Z}.
% Juni 2010

{\bf $\beta$ Pic}: The first direct imaging AO detection with VLT/NACO
of this companion $0.441 \pm 0.008 ^{\prime \prime}$ ($8.57 \pm 0.18$ AU) off
$\beta$ Pic ($19.440 \pm 0.045$ pc, 8 to 20 Myr, A6) was presented
in \citet{2009A&A...493L..21L}, while the 2nd epoch image with common proper motion confirmation
was presented in \citet{2010Sci...329...57L}. This planet detected at $\sim 10$ AU projected separation
was predicted before by \citet{2007A&A...466..389F} at $\sim 12$ AU semi-major axis with
$\sim 2$-5~M$_{\rm Jup}$ to account for the main warp, the two inner belts, and the falling
evaporaing bodies in the $\beta$ Pic debris disk. \citet{2009A&A...493L..21L} estimated the mass of
the detected object to be $\sim 6$-13~M$_{\rm Jup}$ based on uncertain formation models
at the age and distance of the star. A spectrum of the companion could not yet be obtained.
After two imaging epochs, the different positions of the companion
with respect to the star are consistent with orbital motion \citep{2010Sci...329...57L},
but the object can still be a moving background object.
% 2 July 2010

{\bf RXJ 1609} (1RXS~J160929.1-210524): The first direct imaging AO detection with Gemini
of this companion $2.219 \pm 0.006^{\prime \prime}$ ($\sim 311$ AU) off
RXJ1609 ($\sim 145$ pc and 1 to 10 Myr as member of the Sco-Cen T association, K7-M0)
was presented by \citet{2008ApJ...689L.153L}, while the common proper motion confirmation was
published by \citet{2010ApJ...719..497L} and also confirmed by \citet{2011ApJ...726..113I}.
From the JHK Gemini/NIRI spectra (with the AO system Altair), \citet{2008ApJ...689L.153L,2010ApJ...719..497L}
obtained a spectral type (L2-5) and temperature, and with age and distance of the star,
a mass estimate of 6-11~M$_{\rm Jup}$ from uncertain formation models.
% 10 Aug 2010

{\bf PZ Tel}: The first direct imaging AO detections
of this companion $0.3563 \pm 0.0011 ^{\prime \prime}$ ($18.35 \pm 0.99$ AU)
off the star PZ Tel ($51.5 \pm 2.6$ pc, 8 to 20 Myr as member of the $\beta$ Pic
moving group, G9) were obtained with Gemini/NICI by \citet{2010ApJ...720L..82B}
and with VLT/NACO by \citet{2010A&A...523L...1M}.
They both could also confirm common proper motion.
They estimate the mass of PZ Tel B to be 30-42 or 24-40 M$_{\rm Jup}$, respectively,
again possibly below 25 M$_{\rm Jup}$,
so that this companion could be classified as planetary companion imaged directly.
\citet{2010A&A...523L...1M} not only show common proper motion between A and B with $\ge 39~\sigma$,
but also orbital motion of B around A with $\ge 37~\sigma$ including curvature of orbital motion
at $2~\sigma$ significance.
%august 2010

{\bf Ross 458 C}: The imaging detection of the companion C (possibly also to be called {\em b} as
planet candidate) to the
host star binary Ross 458 A+B (12 pc, M0+M7 pair, 150 to 800 Myr) was published in
\citet{2010MNRAS.405.1140G} and \citet{2010A&A...515A..92S} using the 3.5m UKIRT Infrared Deep Sky Survey
together with common proper motion confirmation, with the separation of Ross C
being 102$^{\prime \prime}$ or 1100 AU. The spectroscopic (T8) confirmation was presented in
\citet{2010ApJ...725.1405B} using the 6.5m Magellan/FIRE. The mass of C is estimated to be
$\sim 14$~M$_{\rm Jup}$ from magnitude, temperature, and gravity of the companion
at the distance of the host star \citep{2010ApJ...725.1405B}.
% Dec 2010

{\bf GSC 06214} (GSC 06214-00210): The first direct imaging detections with
the 200-inch Palomar/PHARO and the 10-m Keck/NIRC2 detectors, using a combination
of conventional AO imaging and non-redundant mask interferometry (sparse aperture
mask with AO) of this companion $2.2033 \pm 0.0015 ^{\prime \prime}$ (i.e. $319 \pm 31$ AU)
off the star GSC 06214 (1 to 10 Myr and $145 \pm 14$ pc as member of ScoCen T association, M1)
together with common proper motion confirmation was published by \citet{2011ApJ...726..113I},
who estimate the mass of the companion
from its colors (neglecting extinction, M8-L4), the distance and age of the star,
and cooling models to be $\sim 10$-15~M$_{\rm Jup}$.
%10 Jan 2011

{\bf CD-35~2722}: The first direct imaging AO detections with Gemini-S/NICI
of this companion $3.172 \pm 0.005 ^{\prime \prime}$ (i.e. $67.56 \pm 4.6$ AU)
off the star CD-35~2722 (50 to 150 Myr as member of the AB Dor moving group, $21.3 \pm 1.4$ pc, M1)
together with common proper motion confirmation and spectral classification as L$4 \pm 1$
were published by \citet{2011ApJ...729..139W}, who estimate the mass of the companion
from its luminosity, the age of the star, and cooling models to be $31 \pm 8$~M$_{\rm Jup}$,
or lower when using the temperature of the companion,
so that it could be a planet imaged directly.
As shown by \citet{2011ApJ...729..139W}, the fact that the position angle of this companion
compared to the host star does not change signficantly is inconsistent with a
non-moving background object by $3~\sigma$; the fact that at the same time the separation
between companion and host star does change signficantly, namely exactly according to
what would be expected for a non-moving background object, can either be interpreted as concern
(really co-moving ?) or as evidence for orbital motion: No change in position angle,
but small change in separation would indicate that the orbit is either seen edge-on and/or
strongly eccentric; it is of course also possible that the faint object is a
moving background object or another, but independent, young member of the AB Dor group.
% 10 March 2011

{\bf HIP 78530}: The first direct imaging AO detection with Gemini/NIRI
of this companion $4.529 \pm 0.006 ^{\prime \prime}$ (i.e. $710 \pm 60$ AU)
off HIP 78530 (1 to 10 Myr as member of the ScoCen OB association, B9, $157 \pm 13$ pc)
together with common proper motion confirmation
and spectroscopy (M$8 \pm 1$) were published by \citet{2011ApJ...730...42L},
who determined the mass from formation models to be 19-26~M$_{\rm Jup}$.
HIP 78530 is so far the most massive host star with directly imaged planet (candidate).
%20 March 2011

{\bf WD 0806-661}: The first direct (normal IR) imaging detections with the Spitzer Infrared Array Camera
at 4.5 $\mu $m of this companion $103.2 \pm 0.2 ^{\prime \prime}$ (i.e. $\sim 2500$ AU)
off the White Dwarf WD 0806-661 ($\sim 1.5$ Gyr, $19.2 \pm 0.6$ pc)
together with common proper motion confirmation was published by \citet{2011ApJ...730L...9L}.
From the companion brightness at 4.5 $\mu $m and the rough age of the host star (WD age plus
progenitor life time), \citet{2011ApJ...730L...9L} estimate the mass of the companion to be $\sim 7$~M$_{\rm Jup}$.
\citet{2011ApJ...732L..29R} argue that the host star age (WD plus progenitor) can be as large
as $\sim 2.7$ Gyr, so that the companion mass can be as large as $\sim 13$~M$_{\rm Jup}$,
still in the planetary mass range.
%20 March 2011

{\bf SR 12}: Direct imaging detection with Subaru/CIAO of the companion SR 12 C
(possibly to be called SR 12 b as planetary companion)
$8.661 \pm 0.033 ^{\prime \prime}$ ($1300 \pm 220$~AU) off the close binary SR 12 A+B
(K4+M2.5, $125 \pm 25$ pc and 0.3-10 Myr as member of the $\rho$ Oph star forming cloud)
together with significant common proper motion confirmation (their figure 4 with five epochs)
and a spectrum (M8.5-9.5) were published in \citet{2011AJ....141..119K};
they derived a mass of SR 12 C to be 6-20~M$_{\rm Jup}$.
Given the large separation inside the $\rho$ Oph cloud,
the object C could also be an independant member of the cloud.
Orbital motion or curvature in orbital motion were not yet detected.
% april 2011

{\bf HR 7329}: Direct imaging detection with HST/Nicmos of the companion (one epoch)
$4.194 \pm 0.016 ^{\prime \prime}$ ($200 \pm 16$~AU) off HR 7329 ($\eta$ Tel,
A0, $47.7 \pm 1.5$ pc, 8 to 20 Myr as member of the $\beta$ Pic moving group)
and a spectrum (M7-8) were published in \citet{2000ApJ...541..390L}.
Common proper motion between HR 7329 A and B/b was shown convincingly
(above $3~\sigma$) only recently with new AO imaging \citep{2011MNRAS.tmp.1135N}.
% MN in press

\begin{figure}
\centering
\includegraphics[width=9cm,angle=270]{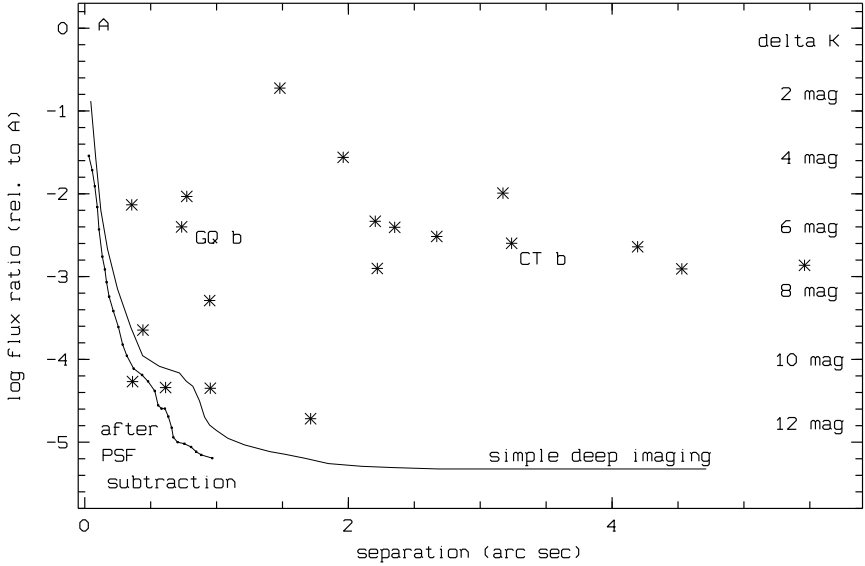}
\caption{We plot the (log of the) flux ratio between the noise level (S/N=3) and the primary host star
(left y axis) or the K-band magnitude difference (right y axis)
versus the separation between companion and host star (in arc sec). The primary host star
is indicated in the upper left by the letter A (log of flux ratio and separation being zero).
The flux ratios of all 20 companions listed in Tables 1 and 2 with separations up to $6 ^{\prime \prime}$
and known K-band flux ratio (known for all but Fomalhaut) are plotted as star symbols
(references for K-band magnitudes for the stars and their companions and
for the separations between them can be found in Sect. 5).
The companions discovered by us (GQ Lup b and CT Cha b) are indicated.
The curve is the dynamic range achieved in our deep AO imaging on GQ Lup
with 102 min total integration time with VLT/NACO; the lower curve with dots
is the dynamic range for the same data achieved after PSF subtraction (figure adapted
from figure 7 in \citet{2008A&A...484..281N}).
All companions above the curve(s) can be detected by this (simple AO imaging) method.
Companions that are fainter and/or closer, i.e. below the curve, cannot be detected.
The PSF subtraction technique can improve the dynamic range and detection
capability at 0.5 to 1$^{\prime \prime}$ by about one magnitude.
The only two companions below the upper dynamic range curve (before PSF subtraction)
are HR 8799 d and e, which were detected by the ADI technique \citep{2008Sci...322.1348M,2010Natur.468.1080M}.
Error bars are omitted for clarity.}
\end{figure}

We notice that the objects GG Tau Bb, TWA 5 B, GJ 417 B \& C, GSC 08047 B/b, LP 261-75 B/b,
HD 203030 B/b, Wolf 940 B/b, G196-3 B/b, PZ Tel B/b, HR 7329 B/b, were not yet listed in
\citet{2011A&A...532A..79S} nor www.exoplanet.eu.
We also note that the object CHXR 73 b \citep{2006ApJ...649..894L}
listed in \citet{2011A&A...532A..79S} and www.exoplanet.eu as planet imaged directly,
is not included in our listing, because common proper motion has not been shown, yet.
A few sub-stellar, possibly planetary mass companions to brown dwarfs, listed as possible
planets in \citet{2011A&A...532A..79S} and www.exoplanet.eu, are also not listed in this
paper, because they are probably not bound, namely
Oph J1622-2405 (also called Oph 1622 or Oph 11, \citet{2006Sci...313.1279J},
UScoCTIO-108 \citep{2008ApJ...673L.185B},
2MASS J04414489+2301513 \citep{2010ApJ...714L..84T}, and
CFBDIRJ1458+1013 AB (also called CFBDS 1458, \citet{2011arXiv1103.0014L}).
2M1207 should therefore also not be listed here; however, we do include it for completness
and comparison, because it is often included in lists of planets imaged directly.

\begin{figure}
\centering
\includegraphics[width=9cm,angle=270]{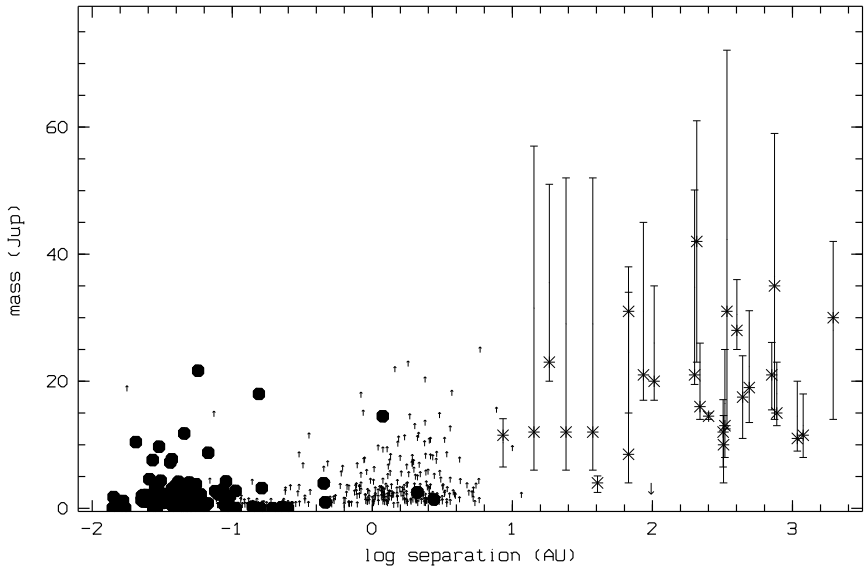}
\caption{We plot the mass of the companion (in Jupiter masses) versus the (log of the)
separation (in AU) for
(i) planet candidates detected by radial velocity only (lower mass limits plotted),
(ii) radial velocity planets confirmed by either astrometry or transit (filled symbols),
where the true masses are known,
and (iii) planets and candidates detected by direct imaging (star symbols)
with masses from the \citet{1997ApJ...491..856B} model as in Tables 2,
because only the Burrows et al. model gives a correct mass for the eSB2 brown dwarf 2M0535 B,
and projected physical separations (calculated from angular separations and distances
as given in Sect. 5); for Fomalhaut b, we plot few M$_{\rm Jup}$ as upper mass limit
at $\sim 100$ AU; only WD 0806-661 is not plotted, because we could not determine its
mass in the same (homogeneous) way as for the others, since the WD 0806-661 companion is
detected only at $4.5~\mu$m, but not in the near-IR.
The four radial velocity planets with the largest separations (8.9 to 11.6 AU)
are $\nu$ Oph c, HIP 5158 c, HIP 70849 b and 47 UMa d.
The directly detected planets (or candidates) with the
smallest projected separations (8.6 to 18.3 AU) are $\beta$ Pic b, HR 8799 e, and PZ Tel b;
these systems are all younger than $\sim 100$ Myr, so that the companions
are still bright enough for direct detection.
This plot shows that the parameter regimes, in which the radial velocity technique
and the direct imaging technique are working, are about to overlap.
This is due to longer monitoring periods for the radial velocity technique
and due to improved technology and, hence, dynamic range in direct imaging.
Hence, it might soon be possible to observe the same planets with
both direct imaging and radial velocity,
which would be best possible for young nearby systems.
The data for radial velocity, transit, and astrometry planets were
taken from www.exoplanet.eu (on 28 July 2011), where the references
for those planets can also be found. We omitted most error bars for clarity,
but we do show the mass errors for the directly imaged planets and candidates.}
\end{figure}

\begin{table}
\begin{tabular}{ccccccc}
\hline
Object    & Burrows 97 & Chabrier 00 & Baraffe 03 & Marley 07 & Baraffe 08 & Wuchterl \\
name      & (L,\,age)  & (L,\,M$_{\rm K}$,\,T,\,t) & (L,\,M$_{\rm K}$,\,T,\,t) & ($\le10$\,Jup) & ($\ge10$\,Myrs)  & (Neuh05)\\
\hline
\multicolumn{7}{c}{Reference object (eSB2 brown dwarf - brown dwarf binary 2M0335):} \\
\hline
2M0535 A      & 50 (45-60)   & 55 (30-60)     & 50 (45-80)     &          &              & 5-13    \\
       B      & 37 (33-46)   & 45 (40-50)     & 43 (40-65)     &          &              & $\le 13$ \\
\hline
\multicolumn{7}{c}{Directly detected planet candidates:} \\
\hline
GG Tau Bb       & 42 (23-61)   & 52 ($\ge 35$)  & 56 ($\ge 41$)  &          &              &   \\
TWA 5 B         & 21 (17-45)   & 23 (20-50)     & 25 (20-50)     &          &              &   \\
GJ 417 B \& C   & 30 (14-42)   & 26 (18-35)     & 25 (20-35)     &          &              &   \\
GSC 08047 B/b   & 16 (14-26)   & 19 (17-25)     & 18 (14-25)     &          &              &   \\
DH Tau B/b      & 13 ( 8-25)   & 20 (6-47)      & 20 (6-50)      & 10 ($\ge 7$)&           & 5 \\
GQ Lup b        & 20 (17-35)   & 25 (20-35)     & 27 (24-37)     &          &              & 1-5 \\
2M1207 b        &  4 (2.5-5)   &  5 (2.5-13)    &  5 (2.5-12)    & 4 (3-5)  &      4       & \\
AB Pic B/b      & 14.5 (14-15) & 16 (14-18)     & 15.5 (11-17)   &          &              & \\
LP 261-75 B/b   & 35 (14-59)   & 26 (16-30)     & 28 (16-32)     &          &              & \\
HD 203030 B/b   & 19 (13.5-31) & 24 (13-28)     & 23 (11-26)     &          &              & \\
HN Peg B/b      & 15 (13-23)   & 21 (14-31)     & 20 (13-27)     & $\ge 10$ &              & \\
CT Cha b        & 17.5 (11-24) & 14 (13-19)     & 16 (13-21)     &          &              & 2-5 \\
Fomalhaut b     & $\le 4.25$   &                & $\le 2$        & $\le 3$  & $\le 2$      & \\
HR 8799 b       & 8.5 (4-38)   & 13 (3-63)      & 12 (4-32)      & 7 ($\ge 3$)& 7 ($\ge 3$)& \\
        c       & 12 (6-52)    & 16 (5-42)      & 12 (6-42)      & 10 ($\ge 6$)& $\ge 5$   & \\
        d       & 12 (6-52)    & 16 (5-42)      & 12 (6-42)      & 10 ($\ge 6$)& $\ge 5$   & \\
        e       & 12 (6-57)    & 16 (5-42)      & 12 (6-42)      & 10 ($\ge 6$)& $\ge 5$   & \\
Wolf 940 B/b    & 28 (25-36)   &                & 33 (24-43)     & \\
G 196-3 B/b     & 31 (12.5-72) & 44 (14-60)     & 43 (11-55)     & \\
$\beta$ Pic b   & 11.5 (6.5-14)& 10 (6-17)      & 9.5 (8-11)     & 10 ($\ge 9$)& 9 ($\ge 8$) & \\
RXJ1609 B/b     & 10 (4-14.5)  &  8 (4-14)      & 8 (4-13)       & 8 ($\ge 4$)&            & \\
PZ Tel B/b      & 23 (20-51)   & 28 (21-41)     & 28 (24-41)     & \\
Ross 458 C      & 11.5 (8-18)  &                & 13 (8-15)      & \\
GSC 06214 B/b   & 12 (6.5-17)  & 15 (6-18)      & 14 (6-23)      & \\
CD-35 2722 B/b  & 31 (15-34)   & 35 (16-43)     & 31 (16-41)     & \\
HIP 78530 B/b   & 21 (15.5-26) & 30 (13-84)     & 32 (11-93)     & \\
WD 0806-661 B/b &              &                & $\le 8.5$      & \\
SR 12 C         & 11 (9-20)    & 10 (5-25)      & 10 (6-25)      & 10 ($\ge 8$)&           & 2-5 \\
HR 7329 B/b     & 21 (19.5-50) & 26 (21-43)     & 27 (23-36)     & \\
%HD 49197 b    & 58 (44-73)   & \\
%DENIS-P J1347-7610 b
%              & 77 (45-93)   & \\
\hline
\end{tabular}
\caption{Masses derived from evolutionary hot-start models.}
\end{table}

\clearpage
\section{Conclusion}

We noticed that the \citet{1997ApJ...491..856B} models give correct masses for
2M0535 B, a young brown dwarf in the eclipsing double-lined spectroscopic
brown dwarf - brown dwarf binary system in Orion, where masses have been
determined without model assumptions \citep{2006Natur.440..311S,2007ApJ...664.1154S}. Hence, we apply
this model for best mass estimates for the planets and candidates imaged directly,
see Fig.~5.

We conclude that direct imaging detection of planets around other stars is possible since
several years with both ground-based AO IR imaging and space-based optical imaging.
For most planets and candidates imaged and confirmed as companions by common proper motion
so far, the planet status is still dubious.
Possibly planetary mass companions apparently co-moving with brown dwarfs,
i.e. apparently forming very wide very low-mass binaries, may well be unbound,
i.e. currently flying apart (\citet{2005AN....326..629M},\citet{2007ApJ...660.1492C}).

Extra-solar planets or candidates as close to their host star as the Solar System planets (within 30 AU)
are still very rare with $\beta$ Pic b, HR 8799 e, PZ Tel B/b, and HR 8799 d being the only exceptions at
8.5, 14.3, 18.3, and 24.2 AU, respectively, all nearby young stars (19 to 52 pc).
As far as angular separation is concerned, the closest planets or candidates imaged directly
are PZ Tel B/b, HR 8799 e, $\beta$ Pic b, HR 8799 d, and GQ Lup b
with separations from 0.36 to 0.75 arc sec.

New AO imaging techniques like ADI, SAM, and locally optimized combination of images
have improved the ability to detect such planets.
Future AO instruments at 8-meter ground-based telescopes will improve the accessible dynamic
range even further. Imaging with a new space based telescope like JWST \citep{2010PASP..122..162B}
or AO imaging at an extremely large telescope of 30 to 40 meters would improve the situation significantly.
Imaging detection of planets with much lower masses, like e.g. Earth-mass planets, might be
possible with a space-based interferometer like Darwin or TPF, but also only around very nearby stars.

{\bf Acknowledgements.} We have used ADS, Simbad, VizieR, WDS, NOMAD, 2MASS,
www.exoplanet.eu, and www.exoplanets.org.
For the image of $\epsilon$ Eri shown in Fig.~1, we would like to thank
Matthias Ammler - von Eiff, who took the observation at VLT (ESO program ID 073.C-0225(A), PI Ammler),
ESO staff at Paranal and Garching for their help with the observations,
and Ronny Errmann, who helped with the data reduction.

\end{document}